\documentclass[aps,twocolumn,pra,tightenlines]{revtex4}
\usepackage{amsmath,amsthm,amsfonts,amssymb,amsbsy,mathrsfs}
\usepackage{units,times,float,graphicx,bm,comment,color}
\usepackage[colorlinks,citecolor=blue,linkcolor=blue,hyperindex,dvipdfm]{hyperref}

\begin{document}
\title{Generating strong mechanical squeezing via combined squeezed vacuum field and two-tone driving}
\author{Xiao-Jie Wu, Huan-Huan Cheng, Cheng-Hua Bai, and Shao-Xiong Wu\footnote{sxwu@nuc.edu.cn}}
\affiliation{School of Semiconductor and Physics, North University of China, Taiyuan 030051, China}
\date{\today}
\begin{abstract}
We propose a novel scheme for generating mechanical squeezed states based on the combined mechanism of a two-tone driving and a squeezed vacuum field. This innovative approach achieves a remarkable improvement in mechanical squeezing performance across the entire range of red/blue detuning ratios. Our study reveals that the squeezed vacuum field not only induces position squeezing of the mechanical oscillator but also facilitates momentum squeezing through phase matching. Moreover, the total squeezing degree exhibits nonlinear enhancement with the increasing of squeezing parameter $r$. The mechanical squeezed state exhibits a $2\pi$-periodic dependence in relation to the squeezing phase $\theta$, offering experimental implementation with a high degree of operational flexibility. Notably, the scheme exhibits strong robustness against cavity dissipation and environmental thermal noise, substantially relaxing the strict parameter-matching requirements inherent in conventional approaches.
\end{abstract}
\maketitle

\section{Introduction}
The seamless integration of quantum optics and optomechanical systems has opened up a new path for the sophisticated manipulation of quantum states in macroscopic mechanical oscillators \cite{Kippenberg2008}. Cavity optomechanics enables the precise quantum-level control of macroscopic objects by exploiting the interaction between optical cavities and radiation pressure \cite{Aspelmeyer2014, Ashkin1980}, thereby addressing fundamental questions within the realm of quantum mechanics. Cavity optomechanics serves as a cornerstone in both the fundamental science and applied technology. In fundamental research, it provides a unique platform for exploring quantum entanglement \cite{D.Vitali2007, Mari2009, DongLH2025, LiJ2021, WeiMR2025, WuSX2023, GuoQ2025, WuSX2024, GuoQ2023, HuCS2020}, mechanical squeezing \cite{WangHF2022, Agarwal2014, JiaoYF2025, BaiCH2020}, photon blockade \cite{GaoXC2023,WuSX2025,WangDY2020,HangR2018,XuXW2024,LinXM2024,ShenHZ2024,XiongW2024}, optomechanically induced transparency \cite{Weis2010,ZhangZM2014,YiXX2018,JingH2019,LaiDG2020}, quantum gravitational effects \cite{Pikovski2012}, and dark matter detection \cite{Manley2021}. Moreover, in applied fields, cavity optomechanical systems enable high-precision weak-force measurements and mass sensing \cite{Fogliano2021}, and also show great promise for quantum information technologies, including the construction of quantum network  \cite{Wallucks2020} and microwave-to-optical transduction \cite{Barzanjeh2022}.

Squeezed states are intrinsically linked to quantum precision measurement \cite{Peano2015}. Consistent with the Heisenberg uncertainty principle, they redistribute quantum noise by suppressing fluctuations in one quadrature component below the standard quantum limit \cite{Walls1983,Scully1997}, with correspondingly amplified fluctuations in the conjugate quadrature component. Recently, we have witnessed substantial advances in the generation and application of mechanical squeezed states, with groundbreaking experimental demonstrations achieving quantum squeezing exceeding 3 dB \cite{Lei2016}. Concurrently, theoretical frameworks have been continuously refined, giving rise to proposed schemes including two-tone driving \cite{Kronwald2013,ZhaoB2024,WuXJ2024,HuangS2021,ZhangW2021}, periodic modulation of external driving amplitudes \cite{HanX2019,GuWJ2013,Guo2023,LiaoJQ2011}, Duffing nonlinearity \cite{LvXY2015,BaiCH2019,XiongBiao2020}, and Kerr effects \cite{Gerry1994,Purdy2013,Garces2016,TongJL2025}.

Squeezed vacuum fields serve as a core nonclassical resource. By suppressing the phase noise of optical fields, they facilitate the more accurate detection of minuscule mechanical displacements, ultimately leading to heightened measurement sensitivity \cite{SunXC2019}. When applied to cavity optomechanical systems, the injection of squeezed vacuum field fundamentally transfers quantum noise suppression to mechanical oscillators via nonlinear optomechanical coupling, significantly enhancing their squeezing performance \cite{Lotfipour2016,Jabri2022,ZhuYJ2024}. In recent years, squeezed vacuum fields have been both theoretically investigated \cite{Walls1983} and experimentally demonstrated \cite{Palomaki2013}. Meanwhile, as a quantum control technique, two-tone driving utilizes two driving fields with distinct frequencies to flexibly engineer a system's effective Hamiltonian. The experimental work in Ref. \cite{Lei2016} achieved 4.7 dB of mechanical squeezing using two-tone driving alone, however it was confined to a specific optimal pump ratio. The scheme proposed in Ref. \cite{WuXJ2024} overcomes the limitation of Ref. \cite{Kronwald2013}, achieving strong mechanical squeezing beyond 3 dB within the sub-optimal red/blue detuning range and enhancing the squeezing strength to 19.49 dB. However, attaining robust mechanical squeezing across the entire operating range remains challenging. Separately, Ref. \cite{Lotfipour2016} demonstrated the transfer of squeezing from an external squeezed vacuum to a mechanics, highlighting another promising path. Based on these prior works, we propose a new scheme that combines a squeezed vacuum field with two-tone driving to utilize their cooperative effects to achieve strong mechanical squeezing in complex dissipative environments. In contrast to the approach in Ref. \cite{Lotfipour2016}, which relies solely on the injected squeezed vacuum field, the synergistic effect in our scheme serves as a key innovation, effectively overcoming the limitations inherent to either method alone and exhibiting remarkable robustness against thermal noise and mechanical dissipation.

Built upon a cavity optomechanical system, this scheme leverages the joint interaction between two-tone driving and squeezed vacuum field injection to jointly achieve the preparation of strongly squeezed mechanical states. The primary aim is to enhance the degree of mechanical squeezing across the range of red/blue detuning ratios through the synergistic effect of these two quantum resources. To further validate and analyze the proposed scheme, in the theoretical computation part, we first analyze the operator dynamics of the optomechanical system using the quantum Langevin equations. Subsequently, by introducing displacement operators, we decompose both the cavity field operator and mechanical displacement operator into their steady-state values and quantum fluctuation components. Following linearization of the Hamiltonian, we perform a frame rotation to obtain the effective Hamiltonian. Finally, we calculate the covariance matrix and introduce squeezing quantification criteria. The total mechanical squeezing degree is characterized by the minimum eigenvalue of the mechanical oscillator's covariance matrix, while the squeezing effects of position/momentum quadrature fluctuations are quantified by their corresponding matrix elements. In numerical simulation part, we first investigate the phase dependence of position/momentum squeezing in the mechanical oscillator. We demonstrate the modulatory effect of squeezed vacuum field injection on the mechanical state pre-squeezed by two-tone driving, and characterize the evolution of total mechanical squeezing with respect to the squeezing phase $\theta$. Furthermore, the periodic squeezing dynamics are visually demonstrated through the evolution of Wigner function  \cite{Weedbrook2012}. Finally, we conduct robustness analysis of the system against environmental perturbations.

The paper is structured as follows. In Sec. \ref{sec2}, we introduce the system model and Hamiltonian, derive the quantum Langevin equations, and obtain the steady-state mean values of both the cavity field and mechanical oscillator. The derivation of the effective Hamiltonian is presented in Sec. \ref{sec3}. In Sec. \ref{sec4}, we solve the covariance matrix and analyze the system's stability. In Sec. \ref{sec5}, we investigate the influence of the squeezing phase $\theta$ and squeezing parameter $r$ on mechanical squeezing, along with robustness analysis of the system. The conclusion is given in the end.

\section{Theoretical model}\label{sec2}
\begin{figure}[t]
\centering
\includegraphics[width=0.75\columnwidth]{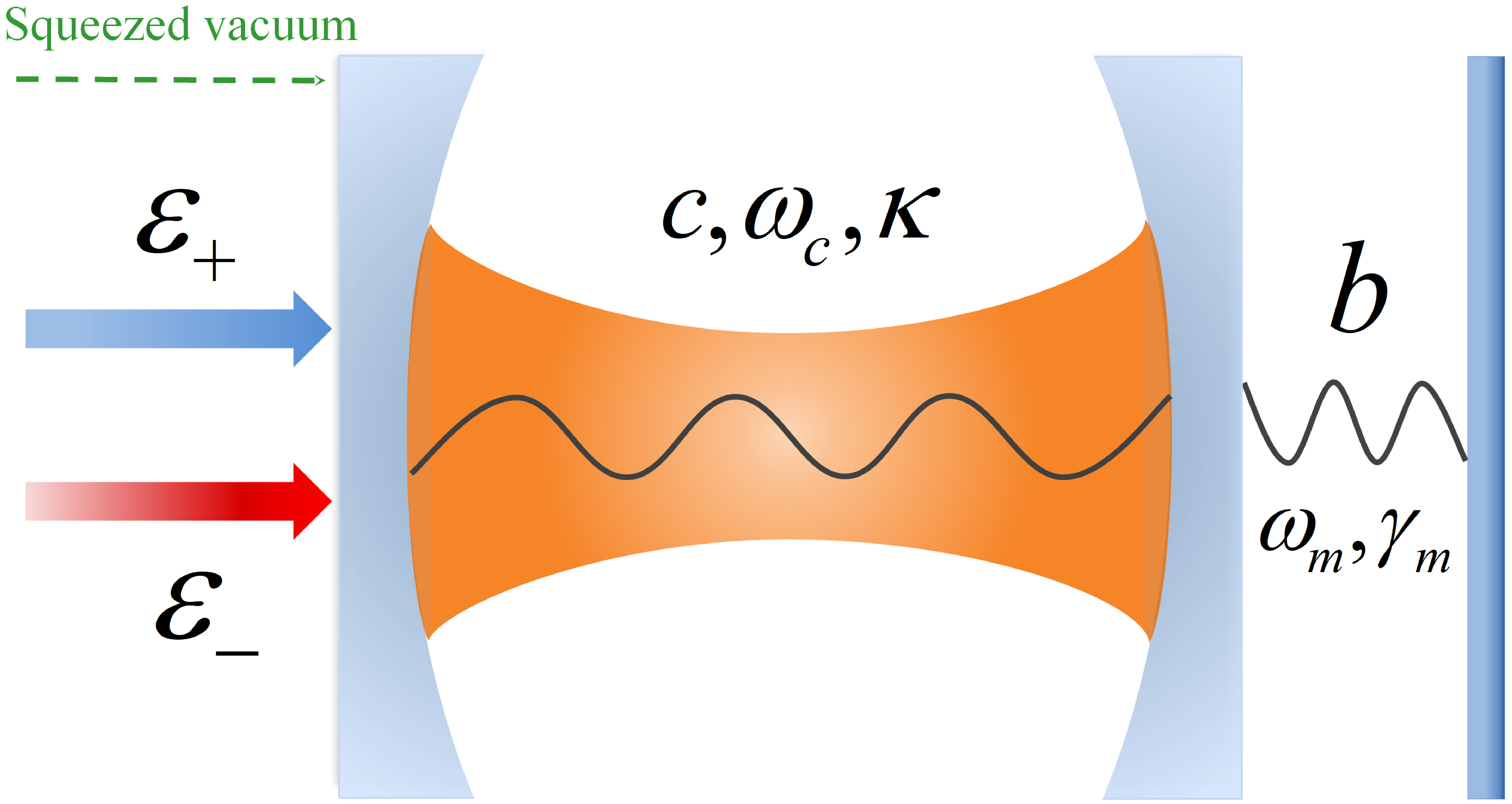}
\caption{Schematic diagram of the system model. The optomechanical system consists of a mechanical oscillator and an optical cavity. The optical cavity is driven by two-tone detuned lasers, one red-detuned and one blue-detuned, and an appropriate squeezed vacuum field is also applied to the optical cavity.}\label{fig:1}
\end{figure}

As shown in Fig. \ref{fig:1}, we consider an optomechanical system consisting of an optical cavity and a mechanical oscillator. By introducing a squeezed vacuum field into the cavity field to suppress quantum noise in the optical field, and simultaneously driving the cavity with red-detuned and blue-detuned lasers, the mechanical oscillator is dynamically coupled to the cavity field. Under this system model, we aim to investigate the squeezing effect on the mechanical resonator. The Hamiltonian of the system is given by (in units of $\hbar$):
\begin{align}
H&=\omega_c{c^\dag}c+\omega_m{b^\dag}b-g_0{c^\dag}c(b^\dag+b)\notag\\
&+[(\varepsilon_+e^{-i\omega_+t}+\varepsilon_-e^{-i\omega_-t})c^\dag+\text{H.c.}]. \label{eq:H}
\end{align}

The first two terms in Eq. (\ref{eq:H}) represent the free Hamiltonians for the cavity field $c$ and the mechanical oscillator $b$, where $\omega_c$ is the frequency of the cavity field with decay rate $\kappa$, and $\omega_m$ is the frequency of the mechanical oscillator with dissipation rate $\gamma_m$. The operators $c$ ($c^\dag$) and $b$ ($b^\dag$) are the annihilation (creation) operators of the cavity field and mechanical oscillator, respectively, which satisfy the commutation relations $[c,{c^\dag}]=1$ and $[b,{b^\dag}]=1$. The third term describes the optomechanical interaction between the cavity field $c$ and the mechanical oscillator $b$ with single-photon optomechanical coupling strength  denoted by $g_0$ \cite{Nunnenkamp2011}. The fourth and fifth terms express the two-tone driving fields applied to the cavity field. Here, $\varepsilon_+$ and $\varepsilon_-$ denote the driving amplitudes for the blue-detuned and red-detuned lasers, respectively, which satisfy the relations $\varepsilon_\pm=\sqrt{\kappa{P_\pm}/\hbar\omega_\pm}$, where $P_+$ and $P_-$ are the corresponding driving powers. The blue-detuned and red-detuned driving frequencies are $\omega_+$ and $\omega_-$, respectively.

Since physical systems unavoidably interact with the surrounding environment, it is essential to take into consideration both dissipative and noise terms when describing their dynamical evolution. The quantum Langevin equations for the mechanical oscillator and optical cavity are given as follows:
\begin{align}
\dot{c}=&-(i\omega_c+\frac{\kappa}{2})c+i{g_0}c(b^\dag+b)-i(\varepsilon_+e^{-i\omega_+t}\notag\\
&+\varepsilon_-e^{-i\omega_-t})+\sqrt{\kappa}c_{\text{in}},\notag\\
\dot{b}=&-(i\omega_m+\frac{\gamma_m}{2})b+i{g_0}{c^\dag}c+\sqrt{\gamma_m}b_{\text{in}}.
\label{eq:QLEs}
\end{align}
Here, the driving frequencies of the red-detuned and blue-detuned lasers satisfy the relation $\omega_\pm=\omega_c\pm\omega_m$. The operator $c_{\text{in}}$ represents the zero-mean input vacuum noise of the cavity field, and $b_{\text{in}}$ denotes the thermal noise operator associated with the mechanical resonator. Under the Markov approximation, the noise correlation functions of the cavity field satisfy the following conditions\cite{Gardiner1986}
\begin{align}
&\langle{c_{\text{in}}^\dag(t)}c_{\text{in}}(t')\rangle=N\delta(t-t'),\notag\\
&\langle{c_{\text{in}}(t)}c_{\text{in}}^\dag(t')\rangle=(N+1)\delta(t-t'),\notag\\
&\langle{c_{\text{in}}(t)}c_{\text{in}}(t')\rangle=M\delta(t-t'),\notag\\
&\langle{c_{\text{in}}^\dag(t)}c_{\text{in}}^\dag(t')\rangle=M^*\delta(t-t')
\label{eq:cguanlian}
\end{align}
with $N=\sinh^2(r)$ and $M=e^{-i\theta}\sinh(r)\cosh(r)$. The noise correlation functions of the mechanical mode $b_{\rm{in}}$ meet:
\begin{align}
&\langle{b_{\text{in}}(t)}b_{\text{in}}^\dag(t')\rangle=(n_m^{\text{th}}+1)\delta(t-t'),\notag\\
&\langle{b_{\text{in}}^\dag(t)}b_{\text{in}}(t')\rangle=n_m^{\text{th}}\delta(t-t').
\end{align}
Here, the variable $n_m^{\text{th}}=[\exp(\hbar\omega_m/k_BT)-1]^{-1}$ represents the average phonon number intimately related to the temperature $T$ of the thermal bath. Additionally, $k_B$ denotes the Boltzmann constant.

The frequency difference between the red-detuned and blue-detuned driving lasers inevitably leads to the steady-state mean value of the cavity field $\langle{c(t)}\rangle$ exhibiting explicit time dependence. To tackle this issue, we decompose $\langle{c}\rangle$ into its positive-frequency (blue-detuned) and negative-frequency (red-detuned) components. By redefining the driving amplitudes $\langle c \rangle_+$ (positive-frequency, blue-detuned) and $\langle c \rangle_-$ (negative-frequency, red-detuned) to separately characterize the strengths of the two-tone driving, the expression of $\langle{c(t)}\rangle$  can be rewritten as:
\begin{align}
\langle{c(t)}\rangle=\langle{c}\rangle_+e^{-i\omega_+t}+\langle{c}\rangle_-e^{-i\omega_-t}.\label{eq:czhengfu}
\end{align}
Through decomposing the cavity field operator into a linear combination of its steady-state mean value and quantum fluctuations, denoted as $c=\langle c\rangle+\delta c$, we are able to extract the evolution equation governing the steady-state mean value in Eq. (\ref{eq:QLEs})
\begin{align}
\langle\dot{c(t)}\rangle=-(i\tilde{\omega}_c+\frac{\kappa}{2})\langle{c(t)}\rangle -i(\varepsilon_+e^{-i\omega_+t}+\varepsilon_-e^{-i\omega_-t})\label{eq:cdot}
\end{align}
with effective frequency  $\tilde\omega_c=\omega_c-g_0(\langle b\rangle^*+\langle b\rangle)$. By substituting Eq. (\ref{eq:czhengfu}) into Eq. (\ref{eq:cdot}) and taking into account the time derivative of Eq. (\ref{eq:czhengfu}), we can further obtain the system of equations governing $\langle{c}\rangle_+$ and $\langle{c}\rangle_-$ as:
\begin{align}
-i\omega_+\langle{c}\rangle_+e^{-i\omega_+t}=& -(i\tilde{\omega}_c+\frac{\kappa}{2})\langle{c}\rangle_+e^{-i\omega_+t}-i\varepsilon_+e^{-i\omega_+t},\notag\\
-i\omega_-\langle{c}\rangle_-e^{-i\omega_-t}=& -(i\tilde{\omega}_c+\frac{\kappa}{2})\langle{c}\rangle_-e^{-i\omega_-t}-i\varepsilon_-e^{-i\omega_-t}.
\end{align}
The analytical expressions for $\langle{c}\rangle_+$ and $\langle{c}\rangle_-$, which are related to the steady-state value $\langle{c(t)}\rangle$ of the cavity field, can be solved as
\begin{align}
\langle{c}\rangle_\pm=\frac{\varepsilon_\pm}{(\omega_\pm-\tilde\omega_c)+i\kappa/2}
\end{align}
within the resolved-sideband regime $\kappa\ll\omega_m$. For the sake of simplifying the analysis without loss of generality, we assume that the steady-state amplitudes $\langle{c}\rangle_\pm$ of the cavity field are real-valued and proportional to the driving strengths $\varepsilon_\pm$. The steady-state amplitude of the mechanical oscillator satisfies the following equations:
\begin{align}
\omega_m\langle{b}\rangle-g_0|\langle{c}\rangle|^2=0.
\end{align}
The dissipation rate of the mechanical oscillator $\gamma_m$ is considerably smaller than that of the cavity field $\kappa$. In other words, $\gamma_m\ll\kappa$. Due to this significant difference, the influence of the mechanical oscillator's dissipation rate on the system's dynamical evolution is negligible. Under this assumption, terms associated with $\gamma_m$ in the dynamical equations can be reasonably neglected.

\section{Hamiltonian Linearization and Effective Hamiltonian}\label{sec3}
Under strong external laser driving, one can perform a standard linearization procedure on Eq. (\ref{eq:QLEs}) by expressing the operators as the sum of their steady-state values and fluctuation terms, i.e., $c=\langle{c}\rangle+\delta{c}, b=\langle{b}\rangle+\delta{b}$. The fluctuations of the cavity field and the mechanical oscillator are governed by
\begin{align}
\delta\dot{c}=&-(i\tilde{\omega}_c+\frac{\kappa}{2})\delta{c} +ig(\delta{b^\dag}+\delta{b})+\sqrt{\kappa}c_{\text{in}},\notag\\
\delta\dot{b}=&-(i\omega_m+\frac{\gamma_m}{2})\delta{b} +i(g^*\delta{c}+g\delta{c^\dag})+\sqrt{\gamma_m}b_{\text{in}},\label{eq:bQLEs}
\end{align}
where the effective coupling is defined as $g=g_0\langle c\rangle$.

According to Eq. (\ref{eq:bQLEs}), one can derive the linearized Hamiltonian
\begin{align}
H_L&=\tilde{\omega}_c\delta{c^\dag}\delta{c}+\omega_m\delta{b^\dag}\delta{b} -g(\delta{c^\dag}\delta{b}+\delta{c^\dag}\delta{b^\dag})\notag\\
&-g^*(\delta{c}\delta{b^\dag}+\delta{c}\delta{b}).
\end{align}
Next, we  carry out a frame rotation on the linearized Hamiltonian, i.e., $H_{\text{eff}}=U^\dag{H_L}U-i{U^\dag}\dot{U}$, where the time-evolution operator $U(t)$ is defined as $U(t)=\exp[-it(\omega_c\delta{c^\dag}\delta{c}+\omega_m\delta{b^\dag}\delta{b})]$. Incorporating the relationships with Eq. (\ref{eq:czhengfu}), the effective Hamiltonian $H_{\text{eff}}$ is expressed as
\begin{align}
H_{\text{eff}}&=-\delta{c^\dag}(g_+\delta{b^\dag}+g_-\delta{b})-\delta{c}(g_+\delta{b}+g_-\delta{b^\dag})\notag\\
&-\delta{c^\dag}(g_+e^{-2i\omega_mt}\delta{b}+g_-e^{2i\omega_mt}\delta{b^\dag})\notag\\
&-\delta{c}(g_+e^{2i\omega_mt}\delta{b^\dag}+g_-e^{-2i\omega_mt}\delta{b^\dag}).\label{eq:Heff}
\end{align}
By employing the effective Hamiltonian $H_{\text{eff}}$, the quantum Langevin equations can be rewritten as
\begin{align}
\delta\dot{c}=&i[f_1(t)\delta{b^\dag}+f_2(t)\delta{b}]-\frac{\kappa}{2}\delta{c}+\sqrt\kappa{c}_{\text{in}},\notag\\
\delta\dot{b}=&i[f_1(t)\delta{c^\dag}+f_3(t)\delta{c}]-\frac{\gamma_m}{2}\delta{b}+\sqrt{\gamma_m}b_{\text{in}}
\label{eq:zuizhongQLEs}
\end{align}
with coefficients are $g_\pm=g_0\langle{c}\rangle_{\pm}$, $f_1(t)=g_++g_-e^{2i\omega_mt}$, $f_2(t)=g_-+g_+e^{-2i\omega_mt}$, and $f_3(t)=g_-+g_+e^{2i\omega_mt}$.

\section{Dynamical equations for the covariance matrix}\label{sec4}
In order to comprehensively investigate the dynamic behavior of the optomechanical system, the quadrature components of the cavity field fluctuation operator, along with the noise operators, are introduced
\begin{align}
\delta{X}=\frac{\delta{c}+\delta{c^\dag}}{\sqrt2},&\
\delta{Y}=\frac{\delta{c}-\delta{c^\dag}}{i\sqrt2},\notag\\
X_{\text{in}}=\frac{c_{\text{in}}+c_{\text{in}}^\dag}{\sqrt2},&\
Y_{\text{in}}=\frac{c_{\text{in}}-c_{\text{in}}^\dag}{i\sqrt2}.
\end{align}
Concurrently, the position and momentum quadrature operators of the mechanical oscillator, as well as the  associated quadrature noise operators, can be formulated as
\begin{align}
\delta{Q}=\frac{\delta{b}+\delta{b^\dag}}{\sqrt2},&\
\delta{P}=\frac{\delta{b}-\delta{b^\dag}}{i\sqrt2},\notag\\
Q_{\text{in}}=\frac{b_{\text{in}}+b_{\text{in}}^\dag}{\sqrt2},&\
P_{\text{in}}=\frac{b_{\text{in}}-b_{\text{in}}^\dag}{i\sqrt2}.
\end{align}
The linearized effective quantum Langevin equation,  denoted as Eq. (\ref{eq:zuizhongQLEs}), can be transformed in the following concise form:
\begin{align}
\dot{\mathcal{U}}(t)=\mathcal{A}(t)\mathcal{U}(t)+\mathcal{N}(t),\label{eq:dotU}
\end{align}
where the quadrature fluctuation operator vector $\mathcal{U}(t)$ and the quadrature fluctuation noise operators $\mathcal{N}(t)$ are defined as
\begin{align}
\mathcal{U}(t)=&[\delta{X},\delta{Y},\delta{Q},\delta{P}]^T,\notag\\
\mathcal{N}(t)=&[\sqrt\kappa{X}_{\text{in}},\sqrt\kappa{Y}_{\text{in}}, \sqrt\gamma_m{Q}_{\text{in}},\sqrt\gamma_m{P}_{\text{in}}]^T,
\end{align}
and the drift matrix $\mathcal{A}(t)$ is a $4\times4$ matrix
\begin{align}
\mathcal{A}(t)= \left[
  \begin{array}{cccc}
  -\frac{\kappa}{2}   & 0                  & -I(f_{12}^+)            & R(f_{12}^-)\\
         0            &-\frac{\kappa}{2}   & R(f_{12}^+)             & I(f_{12}^-)\\
  -I(f_{13}^+)        & R(f_{13}^-)        & -\frac{\gamma_m}{2}     & 0\\
  R(f_{13}^+)         & I(f_{13}^-)        & 0                       & -\frac{\gamma_m}{2}
\end{array}\right].\label{eq:piaoyijuzhen}
\end{align}
The coefficient $f_{jk}^\pm=f_j(t)\pm f_k(t)$ in Eq. (\ref{eq:piaoyijuzhen}) are the complex coupling, and $R(\cdot)$ and $I(\cdot)$ denote the corresponding real and imaginary parts, respectively. In accordance with the Routh-Hurwitz criterion \cite{DeJesus1987}, the system is stable if and only if all the eigenvalues of the drift matrix $\mathcal A (t)$ have negative real parts. Through a rigorous calculation of the stability conditions associated with eigenvalues, the following inequalities are derived:
\begin{align}
\frac{1}{4}(\kappa+\gamma_m)(4g_-^2-4g_+^2+\gamma_m^2+3\kappa\gamma_m+\kappa^2)>0,\notag\\
\frac{1}{16}(\kappa+\gamma_m)^4(4g_-^2-4g_+^2+\kappa\gamma_m)>0,\notag\\
\frac{1}{16}(4g_-^2-4g_+^2+\kappa\gamma_m)^2>0.\label{eq:RH}
\end{align}
It is evident that the sufficient condition for the solvability of the system is $g_+<g_-$. When conducting investigation on the mechanical squeezing, we will impose constraints based on this condition.

Since the dynamical behavior of the optomechanical system is linear and the environmental noise captures  Gaussian feature, the steady-state evolution will converge to a Gaussian state \cite{Weedbrook2012}. Therefore, the fluctuation behavior of the system can be completely described by a $4\times4$ covariance matrix $\mathcal{V}(t)$, whose matrix elements are defined as $\mathcal{V}(t)_{jk}\delta(t-t')=\langle\mathcal{U}_j(t)\mathcal{U}_k(t')+\mathcal{U}_k(t')\mathcal{U}_j(t)\rangle/2$. The dynamical evolution of the covariance matrix $\mathcal{V}(t)$ is determined by:
\begin{align}
\frac{d\mathcal{V}(t)}{dt}=\mathcal{A}(t)\mathcal{V}(t)+\mathcal{V}(t)\mathcal{A}^T(t)+\mathcal{D}(t),
\end{align}
where the elements of diffusion matrix $\mathcal{D}(t)$ are defined as $\mathcal{D}_{jk}\delta(t-t') =\langle\mathcal{N}_j(t)\mathcal{N}_k(t')+ \mathcal{N}_k(t')\mathcal{N}_j(t)\rangle/2$. In the proposed  model, the diffusion matrix is characterized by the direct sum of the cavity field noise matrix and the mechanical oscillator noise matrix, i.e., $\mathcal{D}=\mathcal{D}_a\oplus\mathcal{D}_b$, and can be expressed as
\begin{align}
\mathcal{D}_a=&\frac{\kappa}{2}\left[
{\begin{array}{cc}
     2N+1+M+M^*    &i(M^*-M)\\
     i(M^*-M)      &2N+1-M-M^*
\end{array}} \right],\notag\\
\mathcal{D}_b=&\frac{\gamma_m}{2}\text{diag}[(2n_m^{\rm{th}}+1),(2n_m^{\rm{th}}+1)].
\end{align}

When studying the squeezing feature of a quadrature component (either position or momentum) of the mechanical oscillator, the squeezing degree can be precisely characterized as (in units of dB)
\begin{align}
S(O)=-10\log_{10}\left[\frac{\langle\delta{O(t)}^2\rangle}{\langle\delta{O(t)}^2\rangle_{\text{vac}}}\right].
\end{align}
In the above equation, $O=Q,P$ represent the position and momentum fluctuation operators of the mechanical oscillator, respectively. For a vacuum environment, $\langle\delta O(t)^2\rangle_{\text{vac}}=1/2$. Typically, when the squeezing degree of the mechanical oscillator exceeds 3 dB, it is considered as strong squeezing.

To facilitate a further investigation into the squeezing effects of the mechanical oscillator, one can focus on the reduced covariance matrix of the mechanical oscillator
\begin{align}
\sigma= \left[
  \begin{array}{cc}
  \mathcal{V}_{33}   & \mathcal{V}_{34}\\
  \mathcal{V}_{43}   & \mathcal{V}_{44}
\end{array}\right].\label{eq:sigma}
\end{align}
Let $\lambda$ be the smallest eigenvalue of the reduced covariance matrix $\sigma$, and this eigenvalue determines the total squeezing degree $S$, which is a scalar quantity representing the maximum degree of squeezing attainable in any quadrature component. The definition is given by
\begin{align}
S=-10\log_{10}\lambda.\label{eq:total}
\end{align}

\section{Squeezing of mechanical oscillator}\label{sec5}

\begin{figure}[t]
\centering
\includegraphics[width=0.75\columnwidth]{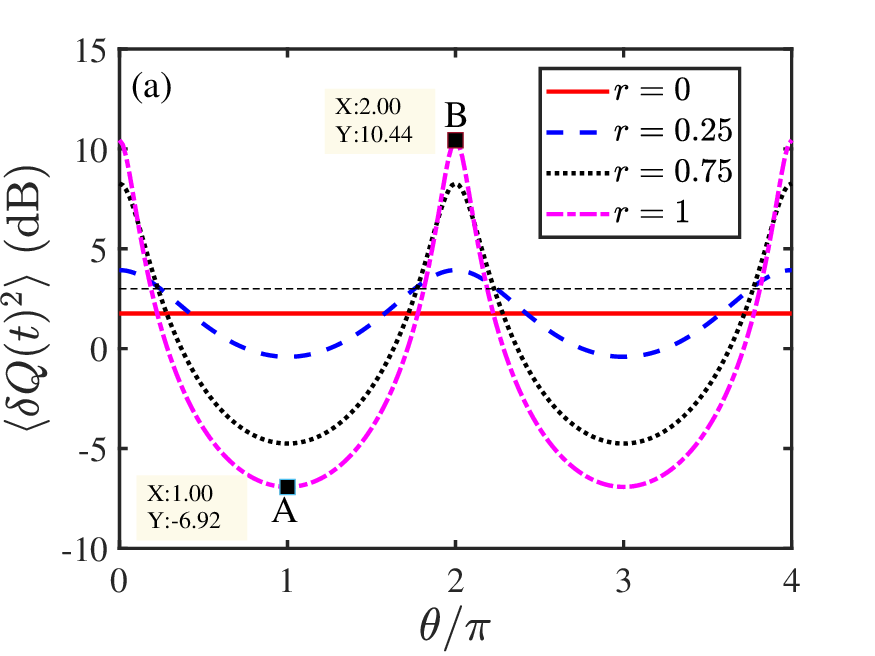}\\
\includegraphics[width=0.75\columnwidth]{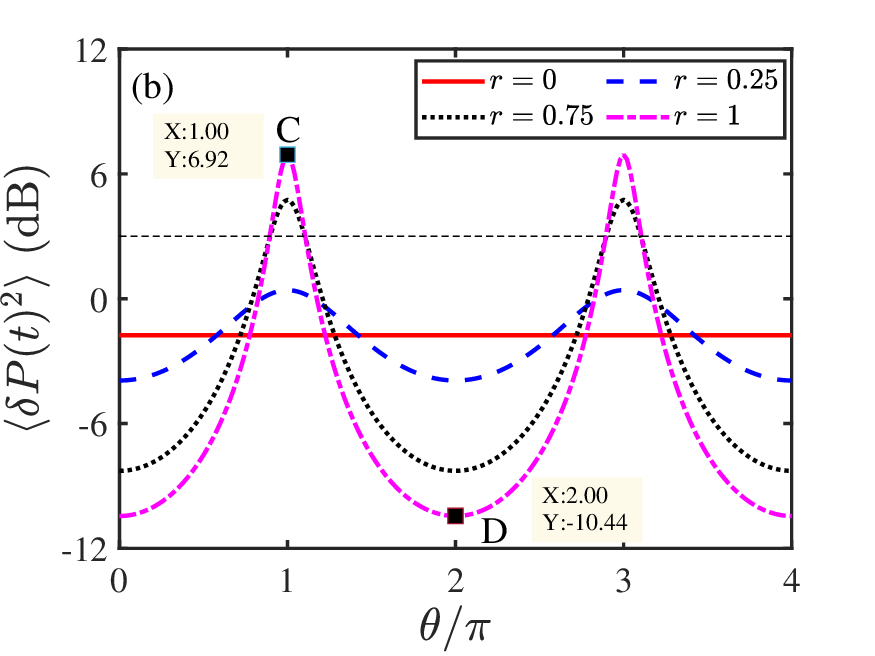}
\caption{Panels (a) and (b) show the position squeezing and momentum squeezing of the mechanical oscillator, respectively, as functions of the phase $\theta$ under different squeezing parameters $r$. The system parameters are set as: $\kappa=0.1\omega_m$, $g_0=10^{-4}\omega_m$, $\gamma_m=10^{-6}\omega_m$, $g_-=0.01\omega_m$, $g_+=0.2g_-$, $n_m^{\text{th}}=0$. The black dashed line indicates the 3 dB squeezing limit.}\label{fig:2}
\end{figure}

Since a squeezed vacuum field is introduced, the related squeezing parameter $r$ and squeezing phase $\theta$ will inevitably affect the squeezing of the mechanical oscillator. Using the mechanical oscillator's frequency $\omega_m$ as the normalization unit, we select the following parameters: $\kappa=0.1\omega_m$, $g_0=10^{-4}\omega_m$, $\gamma_m=10^{-6}\omega_m$, $g_-=0.01\omega_m$. First, we investigate how the phase $\theta$ variations influence the mechanical squeezing under the rotating-wave approximation for different squeezing parameter $r$. In Fig. \ref{fig:2}, we plot how the degree of squeezing for the position and momentum quadrature varies with the phase $\theta$. The results demonstrate that as the phase $\theta$ varies, both the position and momentum quadratures of mechanical oscillator exhibit squeezing effect, and the squeezing degree shows a $2\pi$-periodic change. This observed phenomenon indicates that the squeezing effect is not restricted to a particular quadrature but can alternately manifest along different linear combinations of position and momentum. Specifically, when the phase $\theta=0$ or is an integer multiples of $2\pi$, the mechanical oscillator achieves optimal squeezing in the position quadrature; in contrast, when $\theta=\pi$ or is an odd-integer multiples of $\pi$, optimal squeezing occurs in the momentum quadrature. For both Fig. \ref{fig:2}(a) and Fig. \ref{fig:2}(b), when there is no squeezed vacuum field, i.e., $r=0$, the position quadrature squeezing of the mechanical oscillator remains invariant regardless of phase variations. This phenomenon can be directly and clearly understood from Eq. (\ref{eq:piaoyijuzhen}). The drift matrix $\mathcal{A}(t)$ is independent of both the squeezing parameter $r$ and phase $\theta$, and the influence of squeezed vacuum field is solely manifested in the noise matrix $\mathcal{D}$. Under the parameter conditions, the squeezing degree increases monotonically with the squeezing parameter $r$ increasing.

A comparative analysis of Figs. \ref{fig:2}(a) and \ref{fig:2}(b) reveals a strict symmetric relationship between the position and momentum quadrature squeezing of the mechanical oscillator at a fixed phase $\theta$. For instance, in both panels, the pink dotted lines are used to mark data points for comparison. When $\theta=\pi$, the corresponding position quadrature squeezing (Point A) reaches -6.92 dB, while the momentum quadrature squeezing (Point C) measures 6.92 dB. Conversely, at $\theta=2\pi$, when the position quadrature squeezing peaks at 10.44 dB (Point B), the momentum quadrature squeezing (Point D) drops to -10.44 dB. This phase-dependent behavior reveals the multi-directional nature of quantum squeezing in optomechanical interactions. More significantly, it demonstrates that mechanical squeezing can be achieved not only through conventional two-tone driving schemes but also via transfer mechanism from the cavity's squeezed vacuum state through nonlinear optomechanical coupling, which overcomes the stringent parameter-matching requirements inherent in two-tone driving approaches. Physically, this effect arises because the pre-engineered squeezed vacuum state of the cavity field transfers its quantum noise suppression to the mechanical mode through nonlinear optomechanical coupling.

\begin{figure}[b]
\centering
\includegraphics[width=0.75\columnwidth]{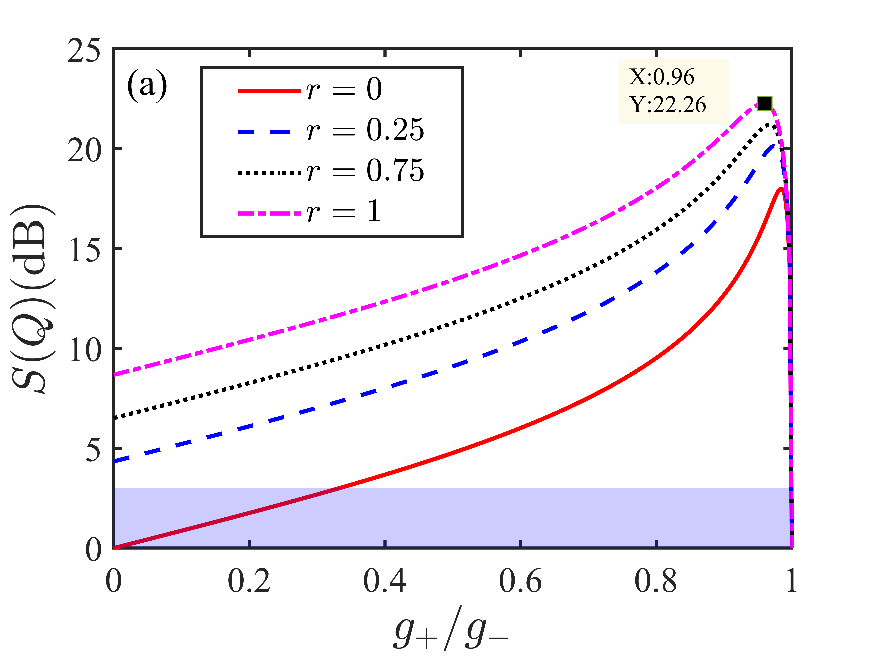}\\
\includegraphics[width=0.75\columnwidth]{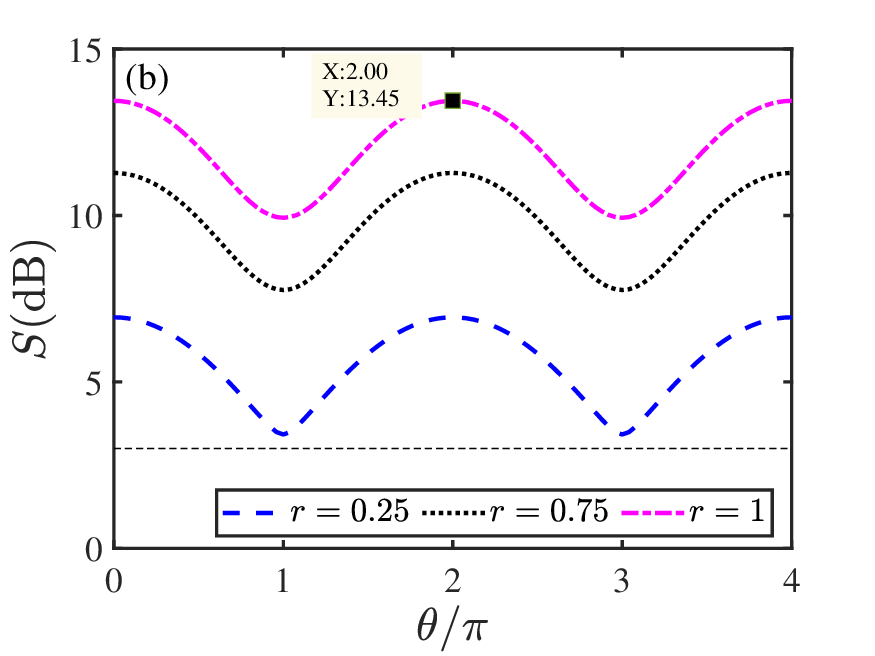}
\caption{Panel (a) shows the position squeezing of the mechanical oscillator as a function of the ratio $g_+/g_-$ under different squeezing parameters $r$. Panel (b) illustrates the dependence of the total squeezing degree $S$ on the phase $\theta$ under different squeezing parameters $r$. The other parameters are the same as those used in Fig. \ref{fig:2}.}\label{fig:3}
\end{figure}

\begin{figure*}
\centering
\includegraphics[width=0.5\columnwidth]{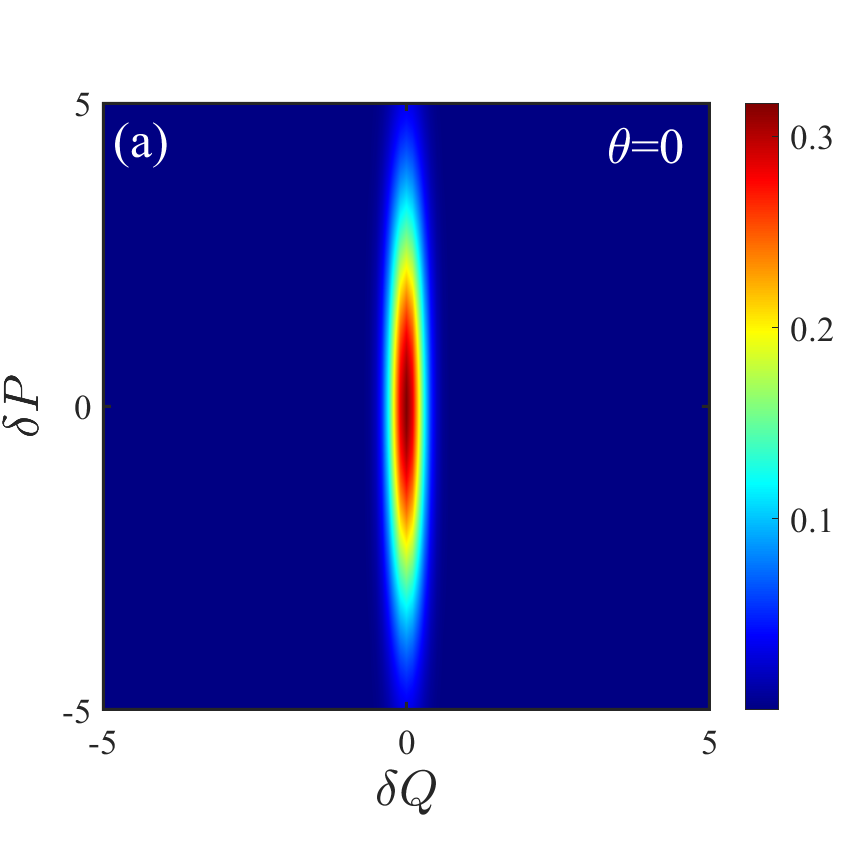}
\includegraphics[width=0.5\columnwidth]{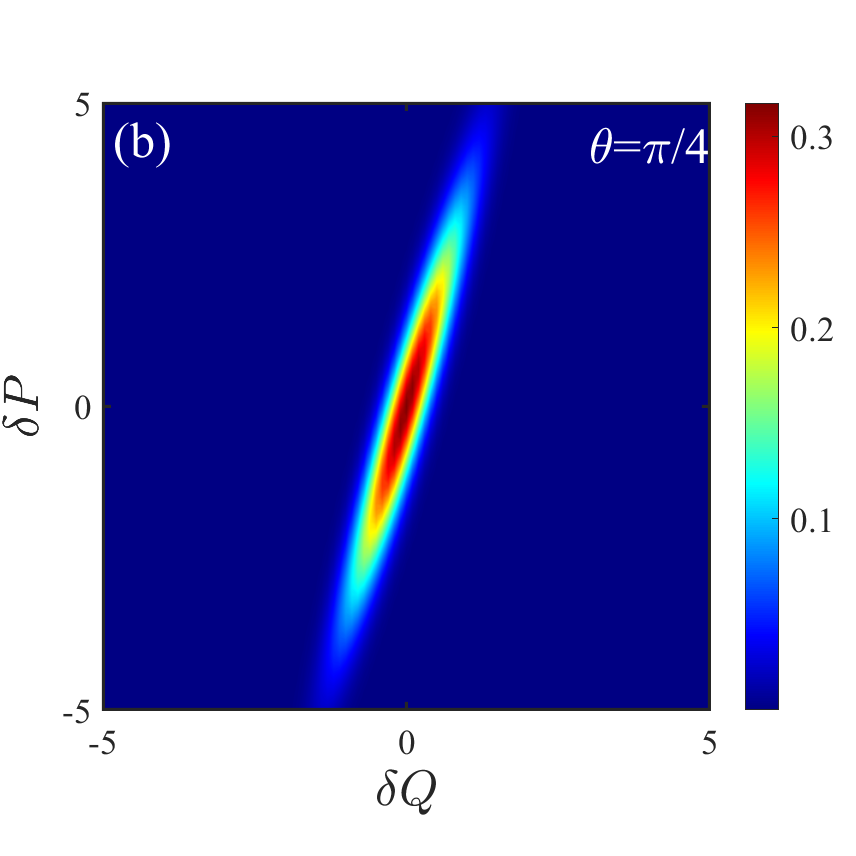}
\includegraphics[width=0.5\columnwidth]{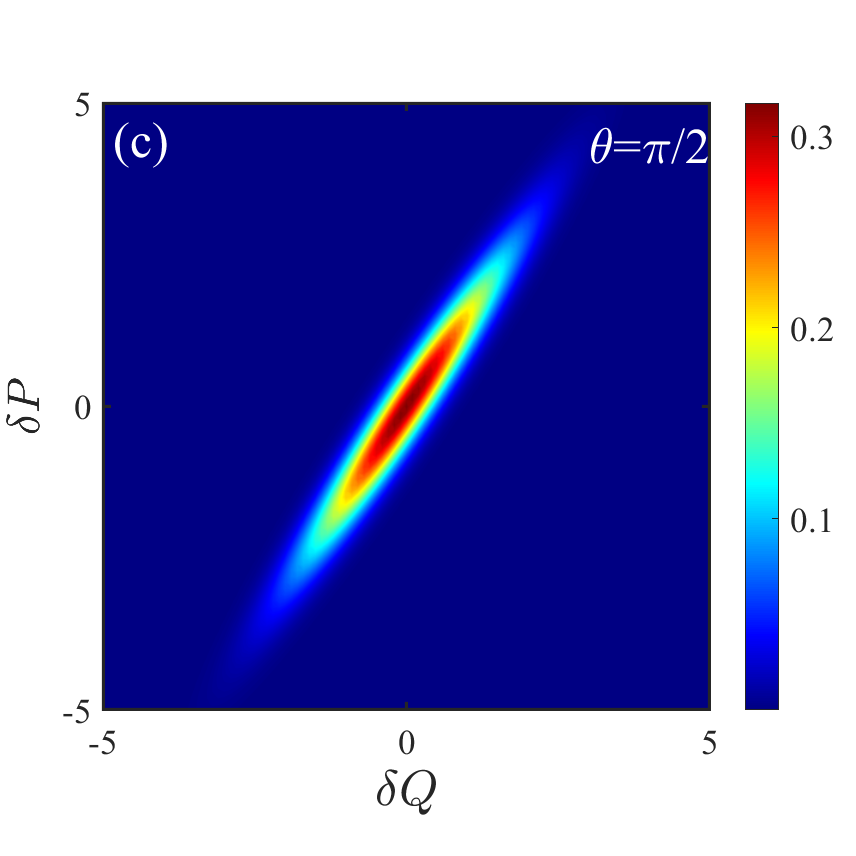}\\
\includegraphics[width=0.5\columnwidth]{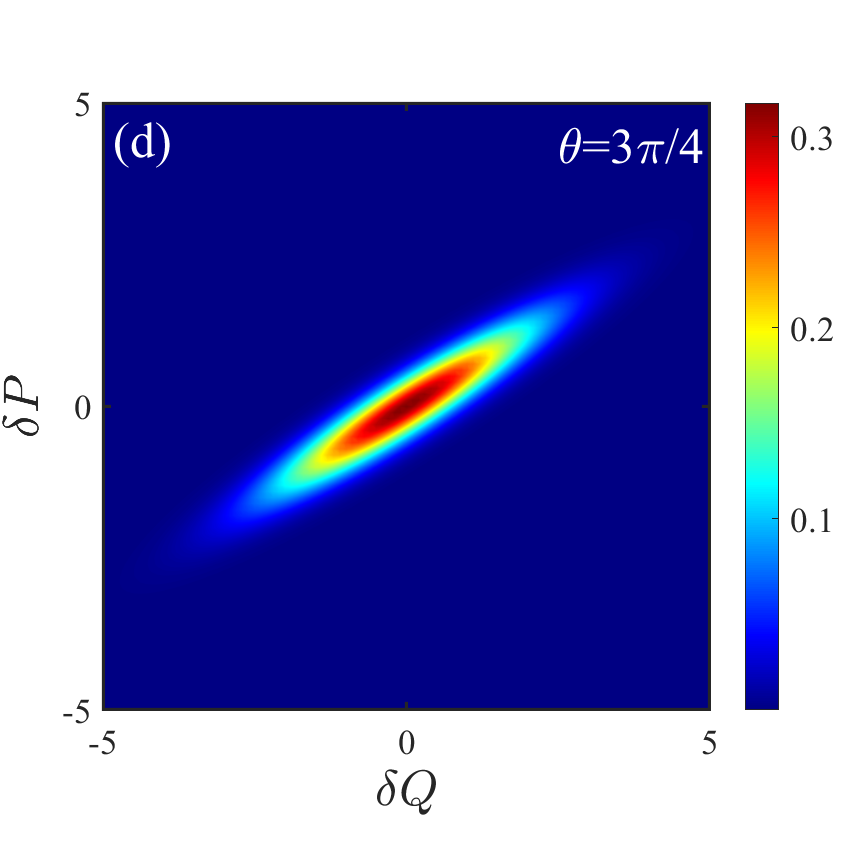}
\includegraphics[width=0.5\columnwidth]{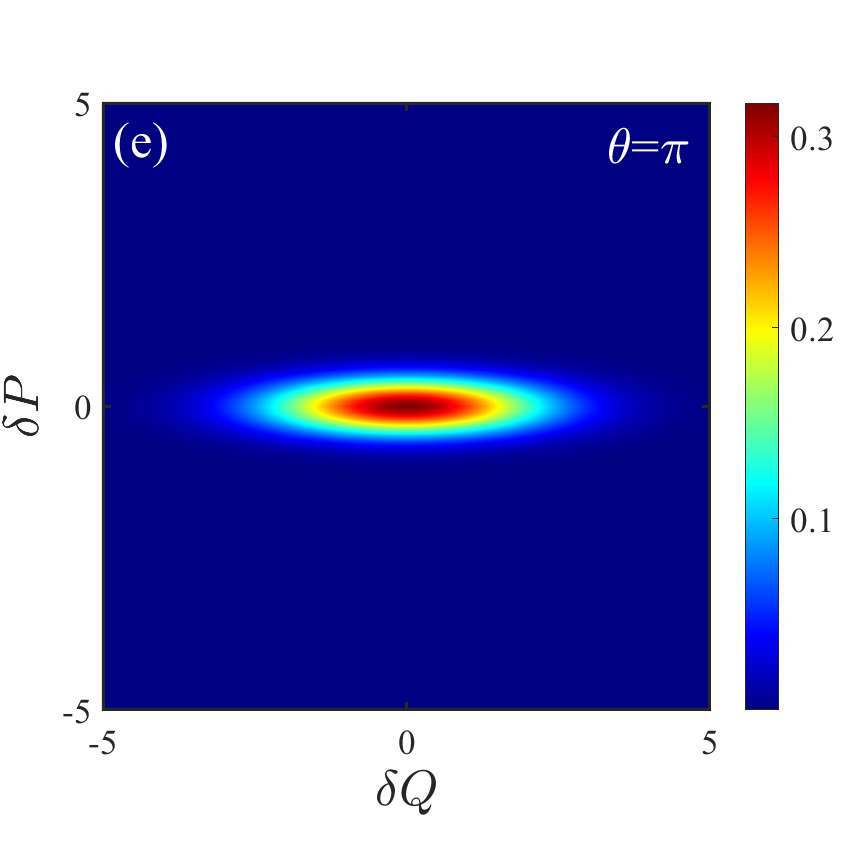}
\includegraphics[width=0.5\columnwidth]{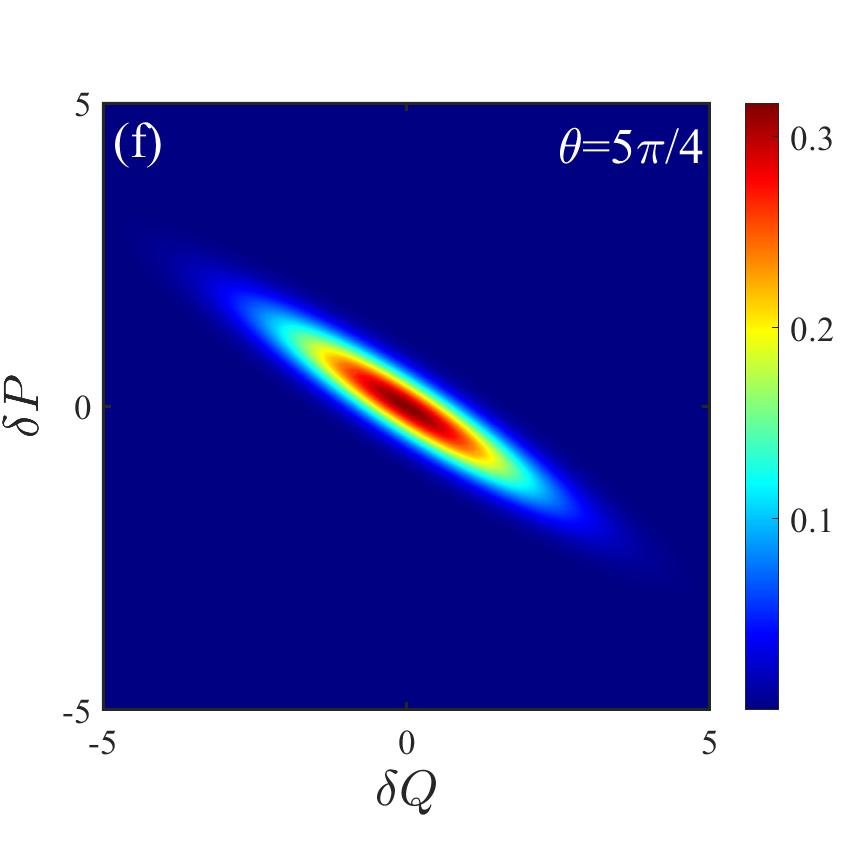}\\
\includegraphics[width=0.5\columnwidth]{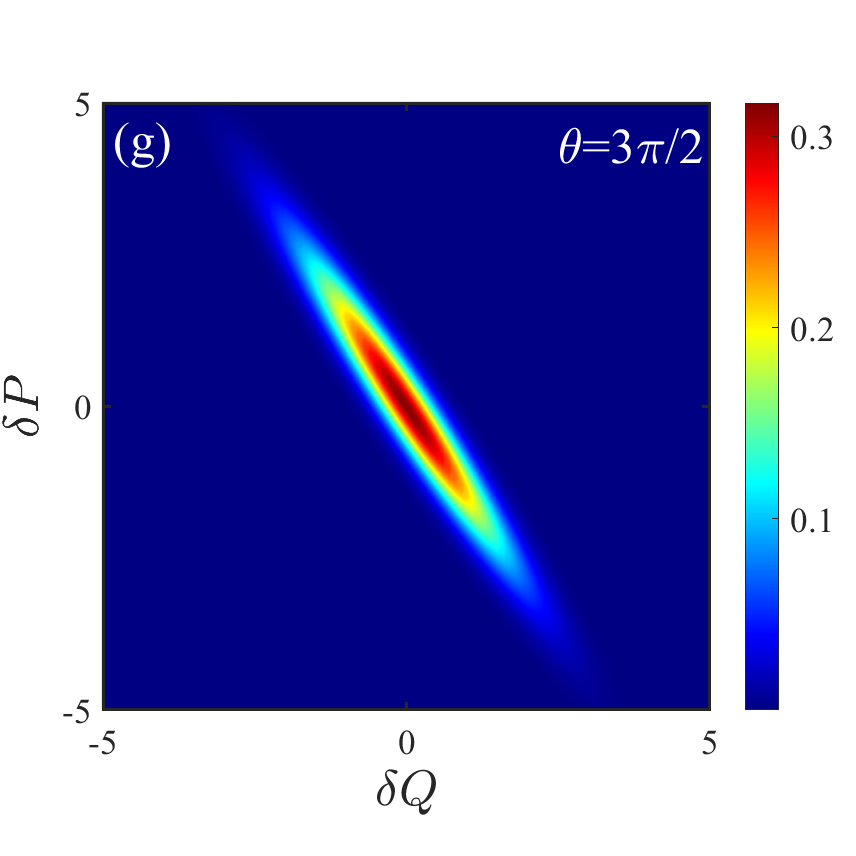}
\includegraphics[width=0.5\columnwidth]{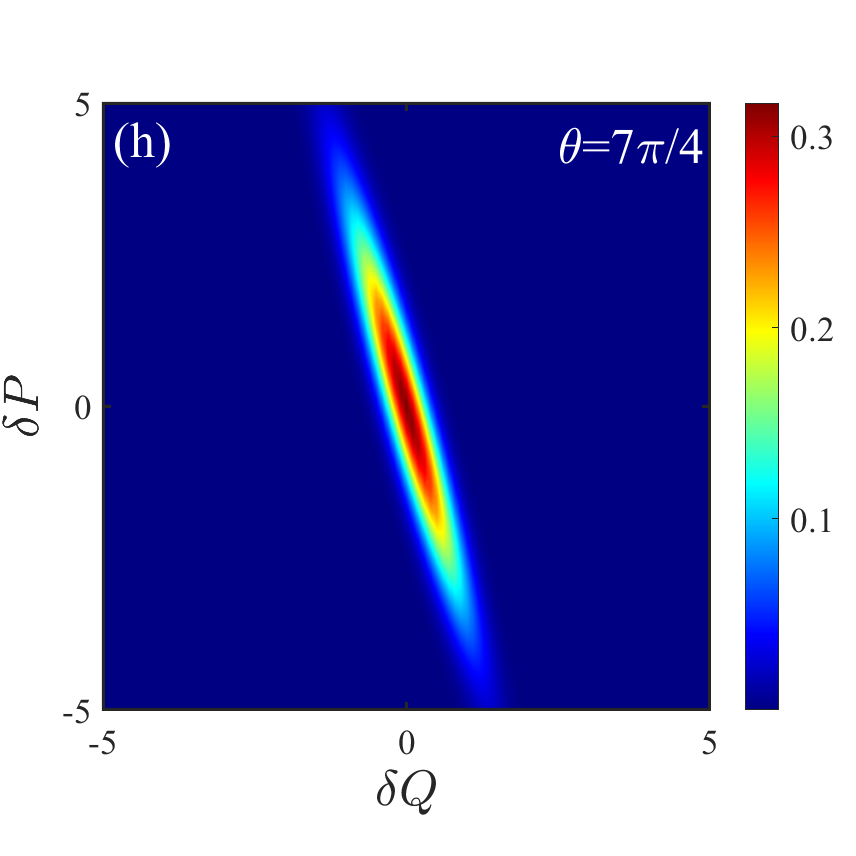}
\includegraphics[width=0.5\columnwidth]{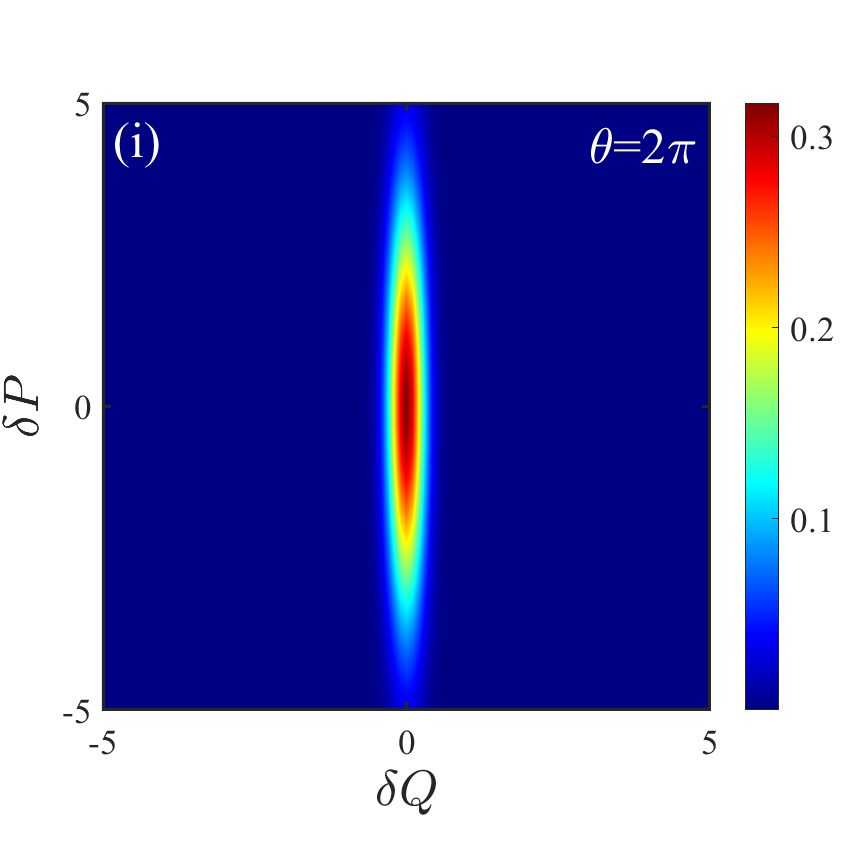}
\caption{The variation of Wigner function of the mechanical oscillator in phase space $\delta{Q}$ and $\delta{P}$ for different squeezing phases $\theta$. The squeezing parameter is set as $r=1$, while the other parameters are the same as those in Fig. \ref{fig:2}.}\label{fig:4}
\end{figure*}

Next, we will further investigate the impact of red/blue detuning ratios on the position squeezing of mechanical oscillator. Numerical simulations presented in Fig. \ref{fig:3}(a) show that as the squeezing parameter $r$ of the squeezed vacuum field increases, the squeezing degree of position is significantly enhanced across the entire red/blue detuning range. Notably, when the red/blue detuning ratio is adjusted to its optimal value, the mechanical squeezing can reach up to 22.26 dB. To further uncover the overall quantum features of the mechanical squeezed state, Fig. \ref{fig:3}(b) illustrates the phase-dependent evolution of the total squeezing degree $S$ in Eq. (\ref{eq:total}), which is quantified by the minimum eigenvalue $\lambda$ of the reduced mechanical covariance matrix $\sigma$ in Eq. (\ref{eq:sigma}). Under conditions where the red/blue detuning ratio $g_+/g_-=0.2$ and the squeezing parameter $r=0$, the position squeezing fails to exceed the 3 dB strong squeezing limit. However, upon the introduction of the squeezed vacuum field and with the consideration of total squeezing degree, strong mechanical squeezing can be readily achieved. Comparison with Fig. \ref{fig:2} reveals that while the position and momentum variances of the mechanical oscillator can individually fall below the standard quantum limit at specific phases, however the total squeezing degree is not a simple linear superposition of these individual variance. For example, when the phase $\theta$ is equal to even multiples of $\pi$, the maximum total squeezing degree $S$ can reach 13.45 dB. This effect originates from the regulatory role of the off-diagonal elements $\mathcal{V}_{34}, \mathcal{V}_{43}$ in the reduced covariance matrix $\sigma$. The quantum nature of this phenomenon cannot be fully captured by measuring only the position squeezing or only the momentum squeezing. Furthermore, the total squeezing degree $S$ exhibits persistent $2\pi$-periodic oscillations with respect to the phase $\theta$, demonstrating effective squeezing across all quadrature directions of the mechanical oscillator. Notably, the amplitude of the mechanical squeezing shows an enhancement as the squeezing parameter $r$ increases.

To visually illustrate the characteristics of phase-space distributions of position and momentum fluctuations (denoted as $\delta{Q}$ and $\delta{P}$) for the mechanical squeezed state, we further display the variation of Wigner function related to the total squeezing degree $S$ in Fig. \ref{fig:4}. The Wigner function is defined as
\begin{align}
\mathcal{W}(\mathcal{R})=&\frac{\exp(-\frac{1}{2}\mathcal{R}^T\sigma^{-1}\mathcal{R})} {2\pi\sqrt{\det(\sigma)}},\label{eq:Wigner}
\end{align}
where $\sigma$ is the $2\times2$ reduced covariance matrix of the mechanical oscillator. The evolution behaviors are plotted at intervals of $\pi/4$ as the phase $\theta$ varies from 0 to $2\pi$. In the Wigner function, the major axis of the ellipse rotates periodically with the phase $\theta$, indicating that the squeezing direction of the mechanical oscillator undergoes regular variations in phase space. When $\theta=0$, the Wigner function is compressed along the position axis (Q-direction) and broadened along the momentum axis (P-direction), achieving the maximum squeezing degree in this cycle. As $\theta$ increases, the squeezing degree of the mechanical oscillator gradually decreases; meanwhile, its squeezing direction rotates clockwise in phase space, shifting toward the momentum axis. At $\theta=\pi$, the squeezing degree reaches its minimum. Subsequently, as $\theta$ continues to increase, the squeezing degree starts to grow again, while the squeezing direction rotates back toward the position axis. When $\theta=2\pi$, the system completes a full periodic evolution, and the entire process exhibits strict mirror symmetry about $\theta=\pi$.

The two-tone driving establishes a squeezed state with a fixed orientation in phase space by cooling the mechanical Bogoliubov mode. The phase $\theta$ of the squeezed vacuum field acts as a control parameter that couples the fluctuations of the mechanical position and momentum through coherent interference, which is clearly reflected in the off-diagonal elements $\mathcal{V}_{34}$ and $\mathcal{V}_{43}$ of the mechanical oscillator's reduced covariance matrix $\sigma$. For a Gaussian state, these two off-diagonal elements are equal, i.e., $\mathcal{V}_{34}=\mathcal{V}_{43}$. By adjusting the phase $\theta$, the values of $\mathcal{V}_{34}$ and $\mathcal{V}_{43}$ will be modified and the principal axis of the squeezing ellipse in phase space will be rotated. Consequently, the non-zero off-diagonal correlation $(\mathcal{V}_{34}=\mathcal{V}_{43}\neq0)$, generated by the combined action of the two-tone driving and the squeezed vacuum field, forms the fundamental physical mechanism that enables momentum squeezing in the system.

From the perspective of physical mechanism, the maximum positional squeezing is achieved when the ratio $g_+/{g_-}$ is close to but not equal to 1. Under steady-state condition, the effective Hamiltonian that describes the interaction between the cavity field and the mechanical oscillator in (\ref{eq:Heff}) is $-\delta{c}^\dag(g_+\delta{b}^\dag+g_-\delta{b})-\delta{c}(g_+\delta{b}+g_-\delta{b}^\dag)$. By introducing the Bogoliubov mode $\delta{B}=\cosh(\xi)\delta{b}+\sinh(\xi)\delta{b}$ with squeezing coefficient $\xi=1/2\cdot\ln[(g_-+g_+)/(g_--g_+)]$, the effective interaction Hamiltonian can be further rewritten as $G_{\text{eff}}(\delta{a}^\dag\delta{B}+\delta{a}\delta{B}^\dag)$, where $G_{\text{eff}}=\sqrt{g_-^2-g_+^2}$ represents the effective coupling strength between the cavity field and the Bogoliubov mode. To achieve maximum squeezing of the mechanical oscillator, the Bogoliubov mode $\delta{B}$ should be cooled to its ground state, and the cooling rate is proportional to coefficient $G_{\text{eff}}$. If $g_+$ becomes excessively large and approaches $g_-$, the cooling rate $G_{\text{eff}}$ will approach zero, leading to the fail of the cooling mechanism. The squeezing coefficient $\xi$ has a direct and positive relationship with the mechanical squeezing degree $S$. It is evident that the coefficient $\xi$ is invalid when $g_+/g_-$ is equal to 1.

\begin{figure}[b]
\centering
\includegraphics[width=0.95\columnwidth]{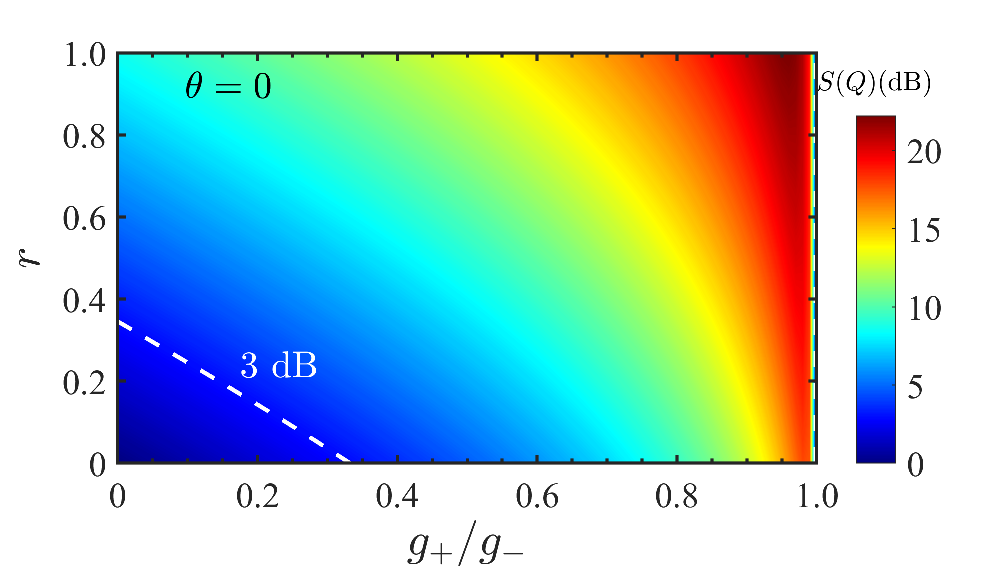}
\caption{The combined effect of the red/blue detuned ratio $g_+/g_-$ and the squeezing parameter $r$ on the squeezing degree of position $S(Q)$. The parameters are the same as those used in Fig. \ref{fig:2}.}\label{fig:5}
\end{figure}

Therefore, to achieve optimal steady-state squeezing, a delicate and balanced interplay between the following two dynamical effects is required: on one hand, $g_+$ should be sufficiently large to achieve a significant value of the squeezing coefficient $\xi$; on the other hand, $g_-$ should remain sufficiently larger than $g_+$ to ensure that the Bogoliubov mode $\delta{B}$ of the mechanical oscillator can be effectively cooled and the optomechanical system remains stable. This optimal balance point occurs when $g_+/g_-$ is close to but slightly less than 1. At this point, the variance of the squeezed quadrature follows the relation $\langle\delta{Q^2}\rangle\propto{e^{-2\xi}}$, which indicates that the variance is exponentially suppressed as $\xi$ increases. When $g_+\approx g_-$, $\xi$ reaches its practical maximum. As a result,  $\langle\delta{Q^2}\rangle$ attains its minimum value, which corresponds to the maximized positional squeezing of the mechanical oscillator.

Fig. \ref{fig:5} illustrates the combined effects of the squeezing parameter $r$ and the red/blue detuning ratio $g_+/g_-$ on the squeezing position quadrature fluctuation in the mechanical oscillator. Based on the preceding analysis, the squeezing phase $\theta=0$ corresponds to the optimal phase for position fluctuation suppression. Under this  optimal phase condition, it can be clearly observed from Fig. \ref{fig:5} that as $r$ and $g_+/g_-$ increase, the  squeezing degree of position $S(Q)$ in the mechanical oscillator gradually enhances. Notably, when appropriate values are selected for both $r$ and $g_+/g_-$, the squeezing degree can easily exceed the  well-known 3 dB strong squeezed limit. In particular, with sub-optimal ranges of the two-tone driving amplitude ratio, the introduction of a squeezed vacuum field allows for the maintenance of high squeezing performance over a wider operational range. This approach effectively reduces the strict dependence on the precise condition of two-tone driving amplitude ratio, which enhances experimental flexibility and fault tolerance, and offers practical advantages for quantum information processing.

\begin{figure}[t]
\centering
\includegraphics[width=0.95\columnwidth]{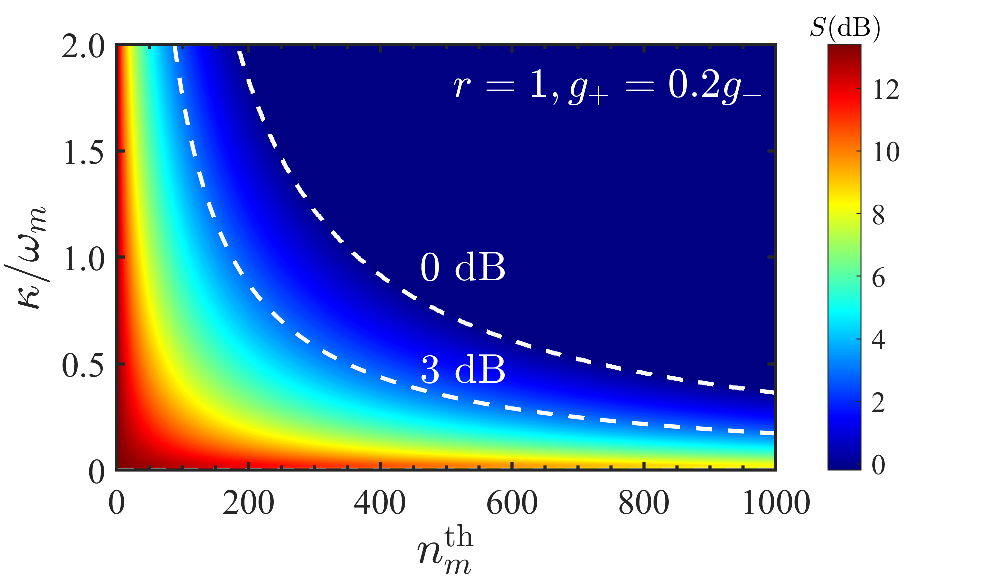}
\caption{The robustness of the total squeezing degree $S$ of the mechanical oscillator with respect to cavity dissipation $\kappa$ and thermal phonon number $n_m^{\text{th}}$ of the mechanical resonator is investigated. The squeezing parameters are set as: $r=1$, and the red/blue detuning ratio is $g_+/g_-=0.2$. The other parameters are the same as those in Fig. \ref{fig:2}.}\label{fig:6}
\end{figure}

Fig. \ref{fig:6} shows the results of numerical simulations that depict how the total squeezing degree $S$ varies as functions of cavity decay rate $\kappa$ and the thermal phonon number $n_m^{\text{th}}$ of mechanical oscillator, aiming to explore the robustness of the optomechanical system in dissipative environments. The results reveal that even under high-dissipation condition $n_m^{\text{th}}=1000$ and $\kappa=2\omega_m$, the system is still capable of achieving significant mechanical squeezing that surpasses the well-known 3 dB threshold. The injection of a squeezed vacuum field into the optomechanical system plays a vital role, and compensates for the noise amplification that is induced by both cavity and mechanical dissipation. It helps to maintain the stability of the squeezed state.

\section{Conclusion}\label{sec6}
In summary, we have constructed a two-mode cavity optomechanical system where a two-tone driving and a squeezed vacuum field act simultaneously on the cavity field, and innovatively studied the quantum squeezing effect of the mechanical oscillator. By thoroughly  investigating the influence of the squeezed vacuum field and the two-tone driving on mechanical squeezing, we found that utilizing the squeezed vacuum field allows for achieving not only position squeezing but also momentum squeezing of the mechanical oscillator. The squeezing phase $\theta$ imparts a periodic variation on the squeezing effect with a period of $2\pi$. An additional significant advantage of this scheme is that the injection of the squeezed vacuum field enables strong mechanical squeezing of the position quadrature within the sub-optimal range of the red/blue detuning ratio. Moreover, it significantly enhances squeezing across the entire parameter region, achieving a squeezing degree of up to 22.26 dB. For the total squeezing degree, it is not a simple linear superposition of position and momentum squeezing; instead, it is intrinsically related to the off-diagonal elements of the reduced covariance matrix. Furthermore, the squeezed vacuum field overcomes the fundamental difficulty in achieving strong mechanical squeezing within non-optimal ranges in the two-tone driving scheme. Notably, the mechanical squeezing exhibits flexible directionality and strict mirror symmetry about $\theta=\pi$. Our study reveals that through the synergistic optimization of the two-tone driving and squeezing parameters, we can significantly enhance the robustness of squeezing, which allows for the maintenance of strong squeezing even under conditions of high dissipation and thermal noise environments. This theoretical framework not only provides a novel approach for realizing strong mechanical squeezed states in optomechanical systems both in theoretical and experimental investigation.

\appendix\label{sec7}
\section{}
We next analyze mechanical squeezing through measurements of the output optical field. By applying the Fourier transformation $\tilde{o}(\omega)=\frac{1}{\sqrt{2\pi}}\int_{-\infty}^\infty{o(t)}e^{i\omega t}dt$ to Eq. (\ref{eq:zuizhongQLEs}), we obtain
\begin{align}
-i\omega\delta\tilde{c}(\omega)=&i[g_-\delta\tilde{b}(\omega)+g_+\delta\tilde{b}^\dag(\omega)] -\frac{\kappa}{2}\delta\tilde{c}(\omega)+\sqrt{\kappa}\tilde{c}_{\text{in}}(\omega),\notag\\
-i\omega\delta\tilde{b}(\omega)=&i[g_+\delta\tilde{c}^\dag(\omega)+g_-\delta\tilde{c}(\omega)] -\frac{\gamma_m}{2}\delta\tilde{b}(\omega)+\sqrt{\gamma_m}\tilde{b}_{\text{in}}(\omega ).
 \label{eq:fuliye}
\end{align}
Following Eq. (\ref{eq:fuliye}), we obtain the approximation for $\delta\tilde{c}(\omega)$:
\begin{align}
\delta\tilde{c}(\omega)=\frac{4i\sqrt{\gamma_m}[g_-\tilde{b}_{\text{in}}(\omega)+g_+\tilde{b}_{\text{in}}^\dag(\omega)] +2\tilde{c}_{\text{in}}(\omega)\sqrt\kappa{v}(\omega)}{d(\omega)},
\end{align}
where $u(\omega)=\gamma_m-2i\omega$, $v(\omega)=\kappa-2i\omega$, $d(\omega)=4g_-^2-4g_+^2+u(\omega)v(\omega)$.

According to the input-output relation of cavity field \cite{Walls1994}
\begin{align}
\delta\tilde{c}_{\text{out}}(\omega)=\sqrt\kappa\delta\tilde{c}(\omega)-\tilde{c}_{\text{in}}(\omega),
\end{align}
can be calculated
\begin{align}
\delta\tilde{c}_{\text{out}}(\omega)=\frac{4i\sqrt{\kappa\gamma_m}[g_-\tilde{b}_{\text{in}}(\omega)+g_+\tilde{b}_{\text{in}}^\dag(\omega)] +[2\kappa{v}(\omega)-d(\omega)]\tilde{c}_{\text{in}}}{d(\omega)}.
\end{align}
Furthermore, output quadrature fluctuations are characterized by
\begin{align}
\delta{Z}_{\text{out}}(\omega)=\frac{1}{\sqrt2}[\delta\tilde{c}_{\text{out}}(\omega)e^{-i\phi} +\delta\tilde{c}_{\text{out}}^\dag(-\omega)e^{i\phi}].
\end{align}
Here, $\phi$ denotes the homodyne measurement phase. For $\phi=0$ and $\phi=\pi/2$, $\delta Z_{\text{out}}(\omega)$ corresponds to amplitude $\delta{X}_{\text{out}}(\omega)$ and phase $\delta{Y}_{\text{out}}(\omega)$ fluctuations, respectively. The fluctuation $\delta{Z}_{\text{out}}(\omega)$ is calculated as
\begin{align}
\delta{Z}_{\text{out}}(\omega)=&A(\omega)X_{\text{in}}(\omega)+B(\omega)Y_{\text{in}}(\omega)\notag\\
&+E(\omega)Q_{\text{in}}(\omega)+F(\omega)P_{\text{in}}(\omega),
\end{align}
where
\begin{align}
A(\omega)=&\frac{2\kappa{v}(\omega)-d(\omega)}{d(\omega)}\cos\phi,\notag\\
B(\omega)=&\frac{2\kappa{v}(\omega)-d(\omega)}{d(\omega)}\sin\phi,\notag\\
E(\omega)=&\frac{4\sqrt{\kappa\gamma_m}}{d(\omega)}\sin\phi(g_++g_-),\notag\\
F(\omega)=&\frac{4\sqrt{\kappa\gamma_m}}{d(\omega)}\cos\phi(g_+-g_-).
\end{align}

The power spectral density of the output field's quadrature fluctuation $\delta{Z}_{\text{out}}(\omega)$ is defined as
\begin{align}
&2\pi{S}_{\text{out}}(\omega)\delta(\omega+\Omega) \notag\\
=&\frac{1}{2}[\langle\delta{Z}_{\text{out}}(\omega)\delta{Z}_{\text{out}}(\Omega)\rangle+ \langle\delta{Z}_{\text{out}}(\Omega)\delta{Z}_{\text{out}}(\omega)\rangle].
\end{align}
Utilizing the input noise correlations in the frequency domain
\begin{align}
\langle{X}_{\text{in}}(\omega)X_{\text{in}}(\Omega)\rangle=&\pi(2N+1+M+M^*)\delta(\omega+\Omega),\notag\\
\langle{Y}_{\text{in}}(\omega)Y_{\text{in}}(\Omega)\rangle=&\pi(2N+1-M-M^*)\delta(\omega+\Omega),\notag\\
\langle{X}_{\text{in}}(\omega)Y_{\text{in}}(\Omega)\rangle=&i\pi(1-M+M^*)\delta(\omega+\Omega),\\
\langle{Y}_{\text{in}}(\omega)X_{\text{in}}(\Omega)\rangle=&i\pi(M^*-M-1)\delta(\omega+\Omega),\notag\\
\langle{Q}_{\text{in}}(\omega)Q_{\text{in}}(\Omega)\rangle=
&\langle{P}_{\text{in}}(\omega)P_{\text{in}}(\Omega)\rangle=2\pi(n_m^{\text{th}}+\frac{1}{2})\delta(\omega+\Omega),\notag\\
\langle{Q}_{\text{in}}(\omega)P_{\text{in}}(\Omega)\rangle=
&-\langle{P}_{\text{in}}(\omega)Q_{\text{in}}(\Omega)\rangle=i\pi\delta(\omega+\Omega).\notag \label{eq:pinyuzhengjiao}
\end{align}

Employing the frequency-domain noise operator correlations from Eq. (\ref{eq:pinyuzhengjiao}), we derive the spectral density of the output field quadrature fluctuations
\begin{align}
S_{Z_{\text{out}}}(\omega)=&\frac{1}{2}A(\omega)A(-\omega)(2N+1+M+M^*)\notag\\
&+\frac{1}{2}B(\omega)B(-\omega)(2N+1-M-M^*)\notag\\
&+\frac{i}{2}A(\omega)B(-\omega)(1-M+M^*)\notag\\
&+\frac{i}{2}B(\omega)A(-\omega)(M^*-M-1)\notag\\
&+[E(\omega)E(-\omega)+F(\omega)F(-\omega)](n_m^{\text{th}}+\frac{1}{2}).
\end{align}
The first four terms originate from vacuum input noise, while the final term arises from mechanical thermal noise. Squeezing of the mechanical resonator is demonstrated where $S_{Z_{\text{out}}}(\omega)<\frac{1}{2}$.

\section*{Acknowledgements}
This work was supported by National Natural Science Foundation of China (Grant No. 12204440).

\end{document}